\documentclass[opre,nonblindrev]{informs3}
\pdfoutput=1
\OneAndAHalfSpacedXI 

\usepackage{endnotes}
\let\footnote=\endnote

\usepackage{natbib}
 \bibpunct[, ]{(}{)}{,}{a}{}{,}%
 \def\bibfont{\small}%

\usepackage[normalem]{ulem}

\TheoremsNumberedThrough     
\ECRepeatTheorems

\EquationsNumberedThrough    

\MANUSCRIPTNO{} 

\usepackage[ruled]{algorithm2e}
\usepackage{verbatim}
\usepackage{enumitem}
\usepackage{bbm}
\usepackage{tikz}
\usetikzlibrary{positioning}
\usepackage{graphicx}
\usepackage{amsmath}
\usepackage{bm}
\usepackage{esvect}
\usepackage{booktabs}

\usepackage{pgfplots}
\pgfplotsset{compat=1.18}
\usepackage[font=footnotesize,labelfont=bf]{caption}

\usepackage{subcaption}
\usepackage{mathrsfs}  
\usepackage{float}
\usepackage{esvect}
\usepackage{amssymb}
\newcommand*{\QEDA}{\null\nobreak\hfill\ensuremath{\blacksquare}}

\usepackage{xr}
\DeclareSymbolFont{extraup}{U}{zavm}{m}{n}
\DeclareMathSymbol{\varheart}{\mathalpha}{extraup}{86}
\DeclareMathSymbol{\vardiamond}{\mathalpha}{extraup}{87}

\usepackage{comment}
\usepackage{color-edits}

\addauthor{ob}{orange}

\begin{document}


\RUNAUTHOR{Besbes, Kanoria and Kumar}

\RUNTITLE{On the Perils of Optimizing the Measurable}

\TITLE{The Fault in Our Recommendations: \\ On the Perils of Optimizing the Measurable}

\ARTICLEAUTHORS{%
\AUTHOR{Omar Besbes}
\AFF{Columbia University, Graduate School of Business, New York, NY, 10027, \EMAIL{ob2105@columbia.edu}} 
\AUTHOR{Yash Kanoria}
\AFF{Columbia University, Graduate School of Business, New York, NY, 10027, \EMAIL{ykanoria@gmail.com}}
\AUTHOR{Akshit Kumar}
\AFF{Columbia University, Graduate School of Business, New York, NY, 10027, \EMAIL{ak4599@columbia.edu}}
}

\ABSTRACT{
    Recommendation systems play a pivotal role on digital platforms by curating content, products and services for users. These systems are widespread, and through customized recommendations, promise to match users with options they will like. To that end,  data on \emph{engagement} is collected and used, with, e.g., measurements of clicks, but also  purchases or consumption times.   Most recommendation systems are ranking-based, where they rank and recommend items based on their predicted engagement. However, the engagement signals  are often only a crude proxy for \emph{user utility}, as data on the latter is rarely collected or available. This paper explores the following critical research question:  {\it By optimizing for measurable proxies, are recommendation systems at risk of significantly under-delivering on user utility? If that is indeed the case, how can one improve utility which is seldom measured?} To study these questions, we introduce a model of repeated user consumption in which, at each interaction, users select between an outside option and the best option from a recommendation set. Our model accounts for user heterogeneity, with the majority preferring ``popular'' content, and a minority favoring ``niche'' content. The system initially lacks knowledge of individual user preferences but can learn these preferences through observations of users' choices over time. Our theoretical and numerical analysis demonstrate that optimizing for engagement signals can lead to significant utility losses. Instead, we propose a utility-aware policy that initially recommends a mix of popular and niche content. We show that such a policy substantially improves  utility despite not measuring it. In fact, in the limit of a forward-looking platform with discount factor $\delta \to 1$, our utility-aware policy achieves the best of both worlds: near-optimal user utility and near-optimal engagement simultaneously. Our study elucidates  an important feature of recommendation systems; given the ability to suggest multiple items, one can perform significant exploration without incurring significant reductions in short term engagement. By recommending high-risk, high-reward items alongside popular items, systems can enhance discovery of high utility items without significantly affecting engagement. 
    
}


\maketitle

\section{Introduction}
Recommendation systems have become integral to numerous online platforms, including social media, e-commerce sites, and media streaming services. These algorithmic mechanisms are deployed by digital platforms to suggest content, products, and services potentially appealing to users. A key promise of these systems is their ability to unveil {\it hidden gems}—content, products, or services users might not have known about but would derive immense {\it utility} from. However, there are questions regarding recommendation systems ability to enable sufficient exploration and the discovery of {\it niche} items which may potentially be of high value to some users. Existing literature \citep{klimashevskaia2023survey} and anecdotal evidence suggests a tendency towards a ``popularity bias'' in contemporary recommendation algorithms, where the algorithms are skewed towards suggesting items that are already popular. The heart of this problem seems to lie in the signals that are measured and optimized for by these recommendation systems. Most recommendations systems use a ranking-based logic to sort and display content by predicted engagement \citep{milli2021optimizing, cunningham2024we}. The signals used to predict and optimize for engagement are generally clicks, likes, comments, and sometimes also include continuous measures of interactions such as dwell time and watch time. While these signals may serve as proxies for {\it utility}, there is potentially a large misalignment between these signals and the {\it utility} generated for the user by the recommended items \citep{ekstrand2016behaviorism}. The fact that a user clicks on a piece of content does not tell us much about how much they {\it value} the content. There has been little concerted effort to measure the {\it user utility} generated by the recommended items. The reason for this is that {\it utility} is quite challenging to measure, and most recommendation systems optimize for signals which can be directly measured, such as clicks, purchases or consumption times. This motivates our key research questions.

\emph{What are the implications of engagement maximization on user utility? Can one improve user utility without measuring it explicitly? And if so, what are the implications on engagement? } 

The broader question of misalignment between engagement and utility maximization by recommendation systems has recently received attention in several papers \citep{chen2021values, milli2021optimizing, stray2021you}; largely through case studies or empirical investigations. We contribute to this growing line of work by studying a stylized model of recommendation systems, allowing us to develop structural insights about the aforementioned misalignment and how it may be addressed.
We study a parsimonious model of a multi-item recommendation system and a repeated content consumption model, where at each interaction, the user selects between an outside option and the best option from a recommended set of items. The recommendation system decides the type(s) of content to recommend at each iteration {given a constraint on the number of items recommended, and 
optimizes for a specified objective. We assume that there are only two types of items -- a popular type $({\sf P})$ which has a fixed positive mean utility and a niche type $({\sf N})$, the utility of which is distributed according to some distribution $F_{\sf N}$ and has a mean utility of zero. Additionally, the user has an outside option which we assume has a mean utility of zero as well. This modelling choice allows us to capture the key tension in recommendation systems -- popular items have similar utility across users whereas niche items can have a high variance across users in terms of their utility. Niche items are not every users' ``cup of tea'' and generally there is a huge variation in the utility generated by the niche items. 
We model niche items as being of low utility for a large fraction of users and being of high utility for a small fraction of users, such that on average the utility is assumed to be zero. In particular, {\it on average across users} the popular type provides more utility than the niche type. For simplicity of exposition, we assume that the platform recommends two items at each iteration.
We now summarize the main contributions of our work.
\begin{enumerate}[label = (\roman*)]
    \item {\bf Theoretical analysis under stylized assumptions}. 
    In Section \ref{sec:main-results}, we will assume that the platform has prior knowledge about the distribution of type utilities, which will be subsequently relaxed in Section \ref{sec:numerical-experiments}. Furthermore, we will assume that the utility of the popular type is fixed and known and the utility of the niche type is a two-point distribution with a single parameter $p$ (cf. \eqref{eq:niche-two-point-distribution}). The parameter $p$ is the fraction of users who derive high utility from the niche type.
    Our analysis in Section \ref{sec:main-results} will focus on the regime where the parameter $p$ is small.
    \begin{enumerate}
        \item {\bf Stark structural misalignment between engagement and utility maximizing policies}.  Our analysis reveals a key structure misalignment between engagement maximizing policies and utility maximizing polices. We show that in the regime of interest $(p \to 0)$, it is engagement maximizing to never recommend the niche types, i.e., the engagement maximizing policy results in homogeneous recommendations, where, at each iteration, it only recommends items of the popular type (cf. Theorem \ref{thm:engagement-maximizing-policy-two-recommendations}). This result, while extreme, is indicative of insufficient exploration and inadequate diversity in many modern day recommendation systems. In contrast, we develop and study a simple utility-aware heuristic {\sf PEAR} (Algorithm \ref{alg:WAH}) which recommends a diverse set of items, and attains significantly higher user utility than the engagement maximizing policy (compare Theorems \ref{thm:engagement-maximizing-policy-two-recommendations} and \ref{thm:utility-maximizing-policy-two-recommednations}). 

    \item {\bf The best of both worlds}. Apart from illuminating the existence of a potentially stark misalignment between engagement and utility, our analysis also characterizes the magnitude of the aforementioned misalignment. Quite strikingly, our analysis uncovers asymmetry in the misalignment between engagement and utility. We observe that by optimizing solely for engagement, there can be substantial loss in the (expected) user utility irrespective of how forward-looking the platform is. 
    In contrast, 
    {our utility-aware heuristic {\sf PEAR} (Algorithm \ref{alg:WAH}) not only achieves approximately optimal utility (Corollary~\ref{cor:asymptotic-optimality-PEAR}) but also near optimal engagement (Corollary \ref{cor:asymmetric-misalignment}), as the platform becomes increasingly forward looking.} 
    Corollary \ref{cor:asymmetric-misalignment} and Figure \ref{fig:asymmetry in misalignment} highlight this asymmetry where we observe a sharp decrease in utility for a minuscule gain in engagement as we move from the utility-aware heuristic {\sf PEAR} to the engagement optimal policy {\sf APP} (defined in \eqref{eq:always-popular-policy}). 
    These findings suggest that platforms are significantly under-delivering on {\it user utility} by optimizing solely for engagement, due to inadequate exploration. Table \ref{tab:utility-engagement-efficiency-with-delta} highlights the magnitude of the misalignment as a function of how forward looking the platform is which is captured by the discount factor $\delta \in [0,1)$. (Small values of $\delta$ correspond to platforms being myopic.) In Table \ref{tab:utility-engagement-efficiency-with-delta}, the mean utility of the popular type is assumed to be $V_{\sf P} = 1$. In the first and second row of the table, we have the relative changes in engagement and utility under the utility-aware heuristic {\sf PEAR} with respect to the engagement optimal policy {\sf APP}, respectively.

    \begin{table}[h]
    \centering
    \caption{Loss in engagement and gain in utility of {\sf PEAR} compared to the engagement-maximizing policy {\sf APP}}
    \begin{tabular}{cccccc}
        \toprule
         & {$\delta = 0$} &  $\delta = 0.9$ & $\delta = 0.99$ & {$\delta = 0.999$} \\
         \midrule
         $\Delta$ Engagement & {$-10.6\%$} & $-1.2\%$ & $-0.12\%$ & {$-0.011\%$} \\
         $\Delta$ Utility & {$+29.3\%$} & $+51.1\%$ & $+53.4\%$ & {$+53.7\%$} \\
         \bottomrule
    \end{tabular}
    
    \label{tab:utility-engagement-efficiency-with-delta}
\end{table}

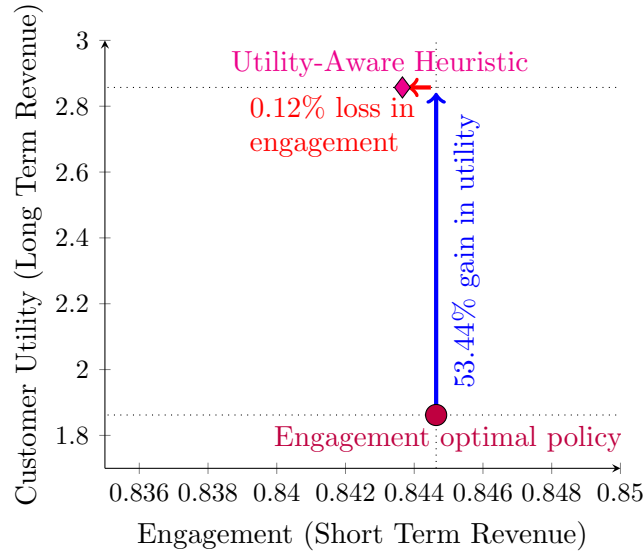
\begin{figure}
    \centering
    \begin{tikzpicture}
        \begin{axis}[
            xlabel={Engagement (Short Term Revenue)},
            ylabel={Customer Utility  (Long Term Revenue)},
            grid=minor,
            xmin=0.835,
            xmax=0.85,
            ymin=1.7,
            ymax=3,
            xticklabel style={
                /pgf/number format/fixed,
                /pgf/number format/precision=3
            },
            scaled x ticks=false,
            enlargelimits=false,
            tick label style={font=\small},
            label style={font=\small},
            enlargelimits=false,
            tick label style={font=\small},
            label style={font=\small},
            axis lines=middle, 
            xlabel style={at={(axis description cs:.5,-0.1)}, anchor=north},
            ylabel style={at={(axis description cs:-0.1,.5)}, rotate=90, anchor=south},
        ]
        \addplot[only marks, mark=*, mark options={fill=purple, mark size = 4pt}] coordinates {
            (0.8446375965030364, 1.861994804058251)
        };
        \node[align=center, text=black, anchor=south] at (axis cs:0.845, 1.72) {\textcolor{purple}{Engagement optimal policy}};
        \draw[-, dotted] (axis cs: 0.8, 1.861994804058251) -- (axis cs: 0.87, 1.861994804058251);
        \draw[-,dotted] (axis cs: 0.8446375965030364, 1.7) -- (axis cs: 0.8446375965030364, 3);
        \addplot[only marks, mark=diamond*, mark options={fill=magenta, mark size = 4pt, draw = none}] coordinates {
            (0.8436564894976852, 2.8570062878811227)
        };
        \node[align=center,  text=black, anchor=north] at (axis cs:0.843, 3) {\textcolor{magenta}{Utility-Aware Heuristic}};
        \draw[-, dotted] (axis cs: 0.8, 2.8570062878811227) -- (axis cs: 0.9, 2.8570062878811227);

        \draw[ultra thick,->,shorten >=2pt,shorten <=2pt, blue] (axis cs:0.8446375965030364, 1.861994804058251) -- (axis cs:0.8446375965030364, 2.8570062878811227);
        \draw[ultra thick, ->, shorten >= 2pt, shorten <= 2pt, red]  (axis cs:0.8446375965030364, 2.8570062878811227) -- (axis cs: 0.8436564894976852, 2.8570062878811227);
        \node [rotate = 90, text width=4cm] at (axis cs: 0.8455, 2.4) { \textcolor{blue}{53.44\% gain in utility}};
        \node [text width=3cm] at (axis cs: 0.8425, 2.74) {\textcolor{red}{0.12\% loss in engagement}};
        

        \end{axis}
    \end{tikzpicture}
    \caption{A comparison of different policies showing that it is possible to improve utility substantially from an engagement-maximizing policy with minimal loss in engagement. Note that the discount factor is $\delta = 0.99$.}
    \label{fig:asymmetry in misalignment}
\end{figure}
    
    \end{enumerate}
    
    \item {\bf Numerical Evidence of generalizability of insights.} In Section \ref{sec:numerical-experiments}, we relax the assumptions that the platform {\it knows} the mean utility of the popular type or the distribution of utilities for the niche type, which is in line with practice where data on possible {\it user utility} values is seldom available. Moreover, instead of assuming a two-point distribution for the utility of the niche type, we consider the class of Pareto distributions with different scale parameters \citep{arnold2008pareto}. 
    Since it is technically challenging to characterize the engagement or welfare optimal policy under the general settings considered above, we will restrict our attention to specific policies -- (i) {\sf Always Popular Policy (APP)} (see \eqref{eq:always-popular-policy}), which as the name suggests, only recommends items of the popular type and 
    (ii) {\sf DIverse-then-CustomizEd (DICE)} policy (Algorithm \ref{alg:ETC}), which recommends a mix of popular and niche type of items for the initial $\mathcal{T}$ periods and switches to recommending either popular or niche type depending on the user's choices in the initial $\mathcal{T}$ periods.   
    Note that neither policy measures or estimates the {\it user utility} from the recommendations. Through a numerical analysis, we observe that as the tail of the Pareto distribution becomes {\it lighter}, the no exploration policy {\sf APP} does strictly better than the exploration driven heuristic {\sf DICE} on engagement, but by a minuscule amount. On the flip side, there is significant utility loss incurred by {\sf APP} vis-a-vis the {\sf DICE} heuristic (cf. Figure \ref{fig:impact-heaviness-tail-delta-0.999}). These results highlight that one can improve substantially upon the engagement maximizing {\sf APP} policy in terms of {\it user utility} -- despite not being able to measure it --  by making diverse and exploratory recommendations, without substantially reducing engagement.
\end{enumerate}

\subsection{Related Literature}
\label{sec:related-literature}

{\it Popularity Bias in Recommendation Systems.} Many present-day recommendation systems are plagued by the ``popularity bias'' where algorithms disproportionately favor already popular items, thereby neglecting niche content that could be valuable to users \citep{ahanger2022popularity,klimashevskaia2023survey, park2008long, celma2008hits}. 
Recommendation systems provide users with a list of items typically ranked in a descending order of predicted engagement \citep{cunningham2024we, adomavicius2005toward}, and the popularity bias is often attributed to this ranking logic \citep{abdollahpouri2017controlling}. 
The topic of popularity bias has also become relevant from a fairness and bias in recommendation systems point of view \citep{ekstrand2022fairness, chen2023bias}.

{\it Measuring Value Beyond Engagement.} The need for more refined measures of engagement has been highlighted in various contexts \citep{besbesetal2016beyond, mcdonald2023impatient, kleinberg2023challenge}.  
A critical challenge in recommendation systems has been measuring the true {\it utility} of recommendations to the users, which extend beyond mere engagement metrics such as clicks and dwell times. Many industrial grade recommendation systems try to incorporate different non-engagement signals as a proxy for {utility} \citep{cunningham2024we, milli2023choosing}. Recent work 
uses a measurement theory approach that constructs a more comprehensive view of value, integrating both observed engagement signals and latent variables to better capture user satisfaction \citep{milli2021optimizing}, however the focus of \cite{milli2021optimizing} is on determining {\it whether} a user values certain content rather than {\it how much} they value it.

{\it Diversity and Novelty in Recommendation Systems.} Instead of solely optimizing for engagement, many recommendations systems have started optimizing for other metrics such as diversity and novelty (refer to \cite{herlocker2004evaluating} for the definition of diversity and novelty in recommendation systems) 
and take a multi-objective optimization approach \citep{milli2023choosing}. There is a growing consensus that diverse recommendations are valuable for users \citep{anderson2020algorithmic, sa2022diversity, steck2018calibrated} and they help in improving long term retention on platforms \citep{chen2021values}. Diverse recommendations also help in covering a user's diverse set of interests and 
mitigate the saturation effects resulting from consuming homogeneous content \citep{ziegler2005improving}. Novelty is another related concept which has been studied in recommendation systems from the perspective of discovery of niche content \citep{castells2021novelty} and this is intimately related to the concept of long tail recommendations \citep{yin2012challenging}. 

\paragraph{{\bf Organization of the paper}} Section \ref{sec:model} describes the model. In Section \ref{sec:main-results} we present the theoretical analysis of a two-point distribution for the base utility of the niche type and use this to discuss the misalignment between engagement and utility. In Section \ref{sec:numerical-experiments}, we demonstrate our key insights under general settings. In Section \ref{sec:conclusion}, we conclude with a discussion of future research directions. For ease of exposition, all the proofs have been deferred to the Appendix.



\section{Model}
\label{sec:model}
We consider an infinite horizon setting with a discount factor $\delta \in [0,1)$. At each time $t \in \mathbb{N}$, the recommendation system recommends $K$ items. For ease of exposition, in this paper we will focus on $K = 2$. There are two {\it types} of items which we refer to as popular ({\sf P}) and niche ({\sf N}) types. There are infinitely many items corresponding to each of these two types. We denote the set of items corresponding to the popular and niche types as $\mathcal{I}_{\sf P}$ and $\mathcal{I}_{\sf N}$ respectively and let $\mathcal{I} = \mathcal{I}_{\sf P} \cup \mathcal{I}_{\sf N}$ be the set of all the items. For a given item $i \in \mathcal{I}$, let $\tau : \mathcal{I} \to \{P,N\}$ denote the type of the item. We denote the set of recommended items at time $t$ as $\pi_t = \{i_{1,t}, i_{2,t}, \dots, i_{K,t}\}$ where $i_{j,t}$ refers to the $j$-th item recommended at time $t$. We denote $\pi = (\pi_0, \pi_1, \pi_2, \dots)$ as the recommendation policy of the platform. The recommendation policy first decides on the type of item to recommend either {\sf P} or {\sf N} and then recommends an item which has not been consumed by the user previously from either $\mathcal{I}_{\sf P}$ or $\mathcal{I}_{\sf N}$. 
Given a set of recommended items $\pi_t$ at time $t$, the user chooses at most one item using an underlying choice model. Let $c_t \triangleq c_t(\pi_t)$ denote the item chosen by the user. Note that we allow for the user to not choose any of the recommended options, in which case we denote the chosen item as $\emptyset$ and refer to $\emptyset$ as the outside option. Each of the two item types, popular and niche, has a base utility, denoted as $V_{\sf P}$ and $V_{\sf N}$ respectively. We assume that the base utility corresponding to the popular product $V_{\sf P}$ is fixed, whereas the base utility corresponding to the niche product $V_{\sf N}$ is a random variable with distribution $F_{\sf N}$. 
We adopt the classical multinomial logit choice model studied in discrete choice literature \cite{anderson1992discrete}:
The utility (or value) that the user derives by consuming item $i$ of type $\tau(i)$ is given as
\begin{align}
    \label{eq:user-utility}
    u_i &= V_{\tau(i)} + \epsilon_i,
\end{align}
where $\epsilon_i$ is an idiosyncratic noise term which is assumed to be a standard Gumbel distribution, i.e. $F_{\epsilon}(x) = \exp(-\exp(-(x - \gamma)))$, where $\gamma$ is the Euler-Mascheroni constant. The noise term $\epsilon$ is zero mean, i.e. $\mathbb{E}[\epsilon] = 0$. The user is a utility maximizer, i.e., given a set of $K$ recommended items $\pi = \{i_1, i_2, \dots, i_K\}$, the user chooses the item with the highest utility
\begin{align}
    \label{eq:user-choice}
    c(\pi) = \argmax_{j \in \pi \cup \{{\emptyset}\}} u_{j} = \argmax_{j \in \pi \cup \{\emptyset\}} V_{\tau(j)} + \epsilon_j,
\end{align}
where $\epsilon_j$ are independent draws from the common distribution $F_\epsilon$. We assume that $V_{\tau({\emptyset})} = 0$, i.e., $u_{\emptyset} \sim F_\epsilon$ and we denote $\tau(\emptyset) \triangleq {\sf O}$. 
Let $\mathcal{H}_t = \{(\pi_0, c_0), (\pi_1, c_0), \dots, (\pi_{t - 1}, c_{t - 1})\}$ denote the history of recommended items and user's choices on the platform up to time $t$. A policy is said to be an online (non-anticipating) policy if the recommendation at time $t$ depends only on the history $\mathcal{H}_t$ up till time $t$. We denote the set of all online policies as $\Pi$.
For any online policy $\pi \in \Pi$, we denote ${\sf Eng}(\pi)$ and ${\sf Util}(\pi)$ as the expected engagement and expected utility achieved under the policy $\pi$.
\begin{align}
    \label{eq:engagement}
    {\sf Eng}(\pi) &= \mathbb{E}\left[\sum_{t = 0}^\infty \delta^t \mathbbm{1}\{c_t(\pi_t) \neq \emptyset\} \right] = \sum_{t = 0}^\infty \delta^t \mathbb{P}(c_t(\pi_t) \neq \emptyset) \\
    \label{eq:utility}
    {\sf Util}(\pi) &= \mathbb{E}\left[ \sum_{t = 0}^{\infty} \delta^t \max_{j \in \pi_t \cup \{\emptyset \} } u_j\right] = \sum_{t = 0}^\infty \delta^t \mathbb{E}\left[\max_{j \in \pi_t \cup \{\emptyset\}} u_j \right]
\end{align}
Note that the expectation is also taken over the distribution of the base utility 
of the niche type.

\section{Analysis of the Two-point distribution for the niche type}
\label{sec:main-results}
In this section, we will focus on a two point distribution for {base utility of} the niche type and assume that this two-point distribution $F_{\sf N}$ is known to the platform. In particular, we will assume that
\begin{align}
    \label{eq:niche-two-point-distribution}
    \mathbb{P}(V_{\sf N} = (1 - p)/ p) = p,  \ \ \ \  \mathbb{P}(V_{\sf N} = -1) = 1 - p
\end{align}
This implies that the mean base utility of the niche type is zero, with a fraction $p$ of users valuing it highly at $V_{\sf N} = (1 - p)/ p$ and the rest valuing it at $V_{\sf N} = -1$. Furthermore, we will assume that the base utility of the popular product $V_{\sf P}$ is positive and known to the platform. 
To start, we study 
{such a} setting where the platform has some knowledge about the base utility of the item types, to highlight the utility loss that occurs when platforms optimize for imperfect proxies of utility such as engagement. We will relax these assumptions in Section \ref{sec:numerical-experiments} where we will not assume knowledge of the base utility of the popular or niche types and we will observe that 
our insights continue to hold in many parametric regimes.

\paragraph{\underline{ {\sf APP}: Engagement Optimal Policy}} Given that on average the popular type generates higher utility than the niche type, a reasonable baseline policy to consider is a ``greedy'' policy, which recommends both items of the popular type, dubbed {\sf Always Popular Policy}, {\sf APP} in short. This policy is intimately related to the ranking-based algorithms typically deployed in practice, where items are  sorted based on their predicted engagement. Since {\it on average across the population}, the popular type generates higher engagement than the niche type due to higher mean base utility, applying the ranking-based logic would result in recommending items only of the popular type. We formally define {\sf APP} as 
\begin{align}
    \label{eq:always-popular-policy}
    {\sf APP} = ((\pi^{\sf APP}_{t})_{t \geq 0}), \ \  \pi^{\sf APP}_{t} = \{i_{1,t}, i_{2,t}\} \text{ s.t. } \tau(i_{j,t}) = {\sf P}, \forall j \in \{1,2\}.
\end{align}
Note that since {\sf APP} never recommends an item of the niche type, it is unable to discern whether or not the niche type is preferred over the popular type by a particular user. This presents us with the classical exploration-exploitation dilemma, where in order to learn whether the niche type generates higher utility {(and engagement)} than the popular type, the platform necessarily needs to explore and recommend items of the niche type, however doing so could potentially hurt engagement. It turns out if the fraction of users $p$ who derive high utility from the niche type is small enough, then {\sf APP} is indeed engagement maximizing. This is formalized below.

\begin{theorem}[Engagement Optimal Policy]
    \label{thm:engagement-maximizing-policy-two-recommendations}
    Fix the base utility of the popular item type $V_{\sf P} \in \mathbb{R}_{+}$ and the discount factor $\delta \in [0,1)$. There exists a $p_0 = p_0(\delta, V_{\sf P}) \in [0,1]$ such that for all $p \leq p_{0}$, then ${\sf APP}$ as defined in \eqref{eq:always-popular-policy} \emph{maximizes} engagement as defined in \eqref{eq:engagement}. Moreover, the expected engagement and utility under {\sf APP} is
    \begin{align*}
        {\sf Eng}({\sf APP}) &= \frac{1}{1 - \delta}\cdot \frac{2e^{V_{\sf P}}}{1 + 2e^{V_{\sf P}}}\, , \\  {\sf Util}({\sf APP}) &= \frac{1}{1 - \delta} \cdot \ln\left(1 + 2e^{V_{\sf P}}\right) .
    \end{align*}
\end{theorem}
The proof of Theorem \ref{thm:engagement-maximizing-policy-two-recommendations} is deferred to Appendix \ref{app:proof-engagement-maximizing-policy-two-recommendations}. Next, we will discuss an exploration-based utility-aware heuristic, and characterize in closed form its  expected engagement and utility.

\paragraph{\underline{{\sf PEAR}: A utility-aware heuristic}} While exploration hurts engagement, it can lead to substantial gains in terms of utility. To this end, we design a simple utility-aware heuristic called 
{\sf Posterior-based Exploration-driver Adaptive Recommendations}, {\sf PEAR} in short, which initially recommends a diverse set of recommendations consisting of one item of the popular type and the other item of the niche type. The utility-aware heuristic {\sf PEAR} maintains a posterior belief on the probability of the niche utility being $(1 - p) / p$ and switches to recommending both items of the popular type when (and if) this posterior belief falls below the initial prior belief $p$. The exploration enables the recommendation system to learn whether the niche type generates high utility for the user or not. We formally describe {\sf PEAR} in Algorithm \ref{alg:WAH}.
\begin{algorithm}
	\SetAlgoNoLine
	\KwIn{Base utility of the popular item $V_{\sf P}$, Parameter $p$ in the niche type distribution (defined in \eqref{eq:niche-two-point-distribution})}
	\SetKwInOut{Initialize}{Initialize}
    \Initialize{$p_0 \gets p$, $S \gets 0$, $F \gets 0$, $\rho_1 \gets \frac{e^{(1 - p) / p}}{1 + e^{V_{\sf P}} + e^{(1 - p) / p}}$, $\rho_2 \gets \frac{e^{-1}}{1 + e^{V_{\sf P}} + e^{-1}}$.}
        \For{$t \in \mathbb{N}$}{
            $\triangleright$ \textcolor{blue}{\small \tt recommend diverse items till posterior at least $p$}\\
            \eIf{$p_{t} \geq p$}{
                $\pi_t = \{(i_1, i_2) : \tau(i_1) = {\sf P} \text{ and } \tau(i_2) = {\sf N} \}$ $\triangleright$ \textcolor{blue}{\small \tt diverse recos} \\
                \eIf{$\tau(c(\pi_t)) = {\sf N}$}{
                    $S \gets S + 1$  $\triangleright$ \textcolor{blue}{\small \tt increment success counter}
                }{
                    $F \gets F + 1$ $\triangleright$ \textcolor{blue}{\small \tt increment failure counter}
                }
            }{
                $\pi_t = \{(i_1, i_2) : \tau(i_1) = {\sf P} \text{ and } \tau(i_2) = {\sf P}\}$  $\triangleright$ \textcolor{blue}{\small \tt only popular recos}
                }{
            }
            $p_{t + 1} = \left(1 + \frac{1 -p}{p} \cdot \frac{\rho_2^{S}(1 - \rho_2)^{F}}{\rho_1^{S}(1 - \rho_1)^{F}} \right)^{-1}$ $\triangleright$ \textcolor{blue}{\small \tt update posterior}
        }
	    
	\caption{{\sf Posterior-based Exploration-driven Adaptive Recommendations (PEAR)}}
	\label{alg:WAH}
\end{algorithm}


\begin{theorem}[Analysis of {\sf PEAR}]
    \label{thm:utility-maximizing-policy-two-recommednations}
    Fix the base utility of the popular item type $V_{\sf P} \in \mathbb{R}_+$ and discount factor $\delta \in [0,1)$. Define $\rho \triangleq 1/(1 + e + e^{V_{\sf P} + 1})$. As $p \to 0$, the expected engagement and utility under {\sf PEAR} (Algorithm \ref{alg:WAH}) is given as 
    \begin{align*}
        {\sf Eng}({\sf PEAR}) &= \frac{1}{1 - \delta \rho} \cdot \frac{e^{V_{\sf P}} + e^{-1}}{1 + e^{V_{\sf P}} + e^{-1}} + \frac{1}{1 - \delta} \cdot \frac{\delta(1 - \rho)}{1 - \delta \rho} \cdot \frac{2e^{V_{\sf P}}}{1 + 2e^{V_{\sf P}}}\, , \\
        {\sf Util}({\sf PEAR}) &= \frac{1}{1 - \delta} + \frac{1}{1 - \delta \rho} \cdot \ln(1 + e^{-1} + e^{V_{\sf P}})  + \frac{1}{1 - \delta} \cdot \frac{\delta(1 - \rho)}{1 - \delta \rho} \cdot \ln(1 + 2e^{V_{\sf P}})\, .
    \end{align*}
\end{theorem}
The proof of Theorem \ref{thm:utility-maximizing-policy-two-recommednations} is deferred to Appendix \ref{app:proof-utility-maximizing-policy-two-recommendations}. 
{Characterizing the utility maximizing policy is technically challenging however we show in Corollary \ref{cor:asymptotic-optimality-PEAR} that {\sf PEAR} is approximately utility optimal for a very forward looking platform ($\delta \to 1$).}
\begin{corollary}[Asymptotic Optimality of {\sf PEAR}]
    \label{cor:asymptotic-optimality-PEAR}
    Let ${\sf Util}({\sf OPT})$ denote the optimal utility obtained by an oracle who at time $t = 0$ knows whether a user prefers niche over popular items or not {(and recommends only the preferred item type)}. Then,
    \begin{align*}
        \lim_{\delta \to 1} \lim_{p \to 0} \frac{{\sf Util}({\sf PEAR})}{{\sf Util}({\sf OPT})} = 1.
    \end{align*}
\end{corollary}

Using Theorems \ref{thm:engagement-maximizing-policy-two-recommendations} and \ref{thm:utility-maximizing-policy-two-recommednations}, it follows that for any discount factor $\delta \in [0,1)$ and $V_{\sf P} > -1$, in the limit $p \to 0$, we have that, 
\begin{align*}
    {\sf Eng}\left({\sf PEAR}\right) < {\sf Eng}\left({\sf APP} \right),  \  \ \ \ {\sf Util}\left( {\sf PEAR}\right) > {\sf Util}\left( {\sf APP} \right).
\end{align*}
The above set of inequalities illuminate the misalignment between engagement and utility. 
Next, we discuss the magnitude of this misalignment, especially in the case when the platforms become increasingly forward looking.

\paragraph{\underline{Asymmetry in the engagement-utility misalignment}} 
As the platforms become increasingly forward looking, i.e., $\delta \to 1$,  we show that engagement under the utility-aware heuristic {\sf PEAR} approaches the optimal engagement achieved by {\sf APP}. In contrast we show that there is a non-vanishing gap between the utility obtained by the utility-aware heuristic {\sf PEAR} and the engagement maximizing policy {\sf APP}. This result is formalized in Corollary \ref{cor:asymmetric-misalignment} below.

\begin{corollary}[Asymmetry in the Misalignment]
    \label{cor:asymmetric-misalignment}
    Fix the attraction parameter of the popular item type $V_{\sf P} \in \mathbb{R}_{+}$. Consider the parameter $p$ in \eqref{eq:niche-two-point-distribution} and the discount factor $\delta$. Then we have that,
    \begin{align}
        \label{eq:asymptotic-engagement-loss}
        \lim_{\delta \to 1} \lim_{p \to 0} \frac{{\sf Eng}({\sf PEAR})}{{\sf Eng}({\sf APP})} &= 1 \\
        \label{eq:asymptotic-utility-gain}
        \lim_{\delta \to 1} \lim_{p \to 0} \frac{{\sf Util}({\sf PEAR})}{{\sf Util}({\sf APP})} &= 1 + \frac{1}{\ln(1 + 2e^{V_{\sf P}})}
    \end{align}
\end{corollary}

Corollary \ref{cor:asymmetric-misalignment} follows from Theorems \ref{thm:engagement-maximizing-policy-two-recommendations} and \ref{thm:utility-maximizing-policy-two-recommednations}. As $\delta \to 1$, from \eqref{eq:asymptotic-utility-gain}, we observe that there is substantial utility gain by the utility-aware heuristic {\sf PEAR} with respect to the engagement maximizing policy {\sf APP}. From \eqref{eq:asymptotic-utility-gain}, it follows that 
\begin{align*}
     \lim_{\delta \to 1} \lim_{p \to 0}\frac{{\sf Util}({\sf PEAR}) - {\sf Util} ({\sf APP})}{{\sf Util}({\sf APP})} \geq \frac{1}{V_{\sf P} + 1} , \ \  \forall \ \  V_{\sf P} \geq \frac{1}{3}
\end{align*}
For $V_{\sf P} = 1$, the above inequality shows that the relative gain in utility {of {\sf PEAR} over {\sf APP}} is approximately $50\%$ which is line with the observations in Figure \ref{fig:asymmetry in misalignment} and Table \ref{tab:utility-engagement-efficiency-with-delta} presented in the introduction. {While characterizing the expected utility and engagement of the utility optimal policy is technical challenging, we expect the utility optimal policy to have similar performance to that of {\sf PEAR} -- small loss in engagement and large gain in utility compared to {\sf APP}}.

Finally, we acknowledge that in this section, we make certain simplifying assumptions such as knowledge of the base utility distribution of the two types, the two-point distribution of the base utility of the niche type and focus on the limiting regime of $p \to 0$. These simplifying assumptions allow us to distill the key structural insights via closed form characterizations, 
with minimal dependence on different parameters of the model. In the following section, we will relax these assumptions and numerically demonstrate that many of our insights continue to hold under varied settings.



\section{Robustness of Insights Under General Settings}
\label{sec:numerical-experiments}
In this section, we will relax the assumptions of Section \ref{sec:main-results}. In particular, we relax the assumption that the mean utility of the popular type and the utility distribution of the niche type is known to the platform. We will restrict our attention to two policies: (i) {\sf APP} as defined in \eqref{eq:always-popular-policy}, and (ii) the {\sf DIverse-then-CustomizEd} ({\sf DICE} in short) heuristic with an exploration parameter $\mathcal{T}$. {\sf DICE} is an explore-then-commit type policy studied in the multi-armed bandit literature \cite{lattimore2020bandit}. The {\sf DICE} heuristic is a {\it prior-free} heuristic which recommends a mix of popular and niche items in the first $\mathcal{T}$ periods and then switches to recommends a customized homogeneous set of recommendation (either popular or niche type) based on the user choices in the first $\mathcal{T}$ periods. We formally describe the {\sf DICE} heuristic in Algorithm \ref{alg:ETC}. We will numerically analyze the case when the {base} utility of the niche type {is drawn from} a generalized Pareto distribution with parameters $\mu, \sigma$ and $\xi$ and denoted as $\text{GDP}(\mu, \sigma, \xi)$. We will focus on a special class of these distributions where we fix $\mu = -1$ and $\sigma = 1 - \xi$ and we only have a single parameter $\xi$; we will denote these distributions as $F^{\xi}_{\sf N}$. For any $\xi < 1$, the mean of the distributions is zero (same as the outside option). The CDF of the distributions as a function of the parameter $\xi$ is provided below.
\begin{align*}
    F_{\sf N}^{\xi}(x) = \begin{cases}
        1 - \left(1 + \frac{\xi}{1 - \xi} \cdot (x + 1) \right)^{-1/\xi},  \ \ \ \ & \xi \neq 0 \\
        1 - \exp\left(- (x+1)\right), \ \ \ \ & \xi = 0
    \end{cases}
\end{align*}
\begin{algorithm}[h]
	\SetAlgoNoLine
	\KwIn{Exploration Phase Length $\mathcal{T}$}
	\SetKwInOut{Initialize}{Initialize}
    \Initialize{${\sf Count}_{\sf P} \gets 0$, ${\sf Count}_{\sf N} \gets 0$}
        \For{$t \in \{1, 2, \dots, \mathcal{T} \}$}{
            $\pi_t = \{(i_1, i_2) : \tau(i_1) = {\sf P} \text{ and } \tau(i_2) = {\sf N} \}$ $\triangleright$ \textcolor{blue}{\small \tt diverse recos} \\
            \uIf{$\tau(c(\pi_t)) = {\sf P}$}{
                ${\sf Count}_{\sf P} \gets {\sf Count}_{\sf P} + 1$ $\triangleright$ \textcolor{blue}{\small \tt increment popular counter}
            }
            \uElseIf{$\tau(c(\pi_t)) = {\sf N}$}{
                ${\sf Count}_{\sf N} \gets {\sf Count}_{\sf N} + 1$ $\triangleright$ \textcolor{blue}{\small \tt increment niche counter}
            }
            \Else{
                continue
            }
        }
        $\triangleright$ \textcolor{blue}{\small \tt check if user chooses more of popular or niche type} \\
        \eIf{
            ${\sf Count}_{\sf P} \geq {\sf Count}_{\sf N}$
        }{
            {\sf PrefType} $= {\sf P}$
        }{
            {\sf PrefType} $= {\sf N}$
        }
        $\triangleright$ \textcolor{blue}{\small \tt recommend the more chosen type during exploratory phase $\{1, 2,\dots, \mathcal{T}\}$}\\
        \For{$t \in \{\mathcal{T} + 1, \mathcal{T} + 2, \dots \}$}{
            $\pi_t = \{(i_1, i_2) : \tau(i_1) = \tau(i_2) = {\sf PrefType}\}$ $\triangleright$ \textcolor{blue}{\small \tt customized recos based on users choices in exploratory phase}
        }
	\caption{{\sf DIverse-then-CustomizEd (DICE)}}
	\label{alg:ETC}
\end{algorithm}


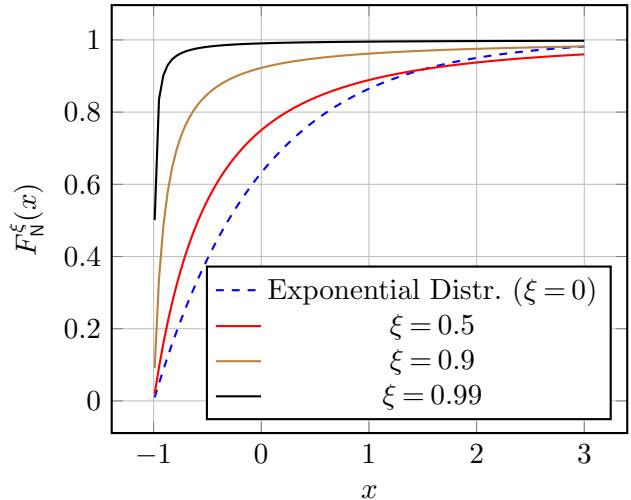
\begin{figure}[h]
    \centering
    \begin{tikzpicture}
\begin{axis}[
    title={},
    xlabel={$x$},
    ylabel={$F_{\sf N}^{\xi}(x)$},
    legend pos=south east,
    grid=major,
    domain=-0.99:3,
    samples=100,
    no marks,
    thick
]
\addplot[blue, dashed] {1 - exp(- (x + 1) / (1 - 0))};
\addlegendentry{Exponential Distr. ($\xi = 0$)}

\addplot[red] {1 - (1 + 0.5 * ((x + 1) / (1 - 0.5)))^(-2)};
\addlegendentry{$\xi = 0.5$}

\addplot[brown] {1 - (1 + 0.9 * ((x + 1) / (1 - 0.9)))^(-1.111)};
\addlegendentry{$\xi = 0.9$}

\addplot[black] {1 - (1 + 0.99 * ((x + 1) / (1 - 0.99)))^(-1.0101)};
\addlegendentry{$\xi = 0.99$}

\end{axis}
\end{tikzpicture}
    \caption{CDF of Generalized Pareto Distribution for $\mu = -1$ and $\sigma = 1 - \xi$}
    \label{fig:gpd}
\end{figure}
For $\xi = 0$, we have the exponential distribution and for $\xi > 0$, we have the Pareto distribution with scale parameter $\xi$, which is a heavy-tailed distribution. 
From Figure \ref{fig:gpd}, we observe that as $\xi$ increases, the probability of the niche type having positive {base} utility decreases. 
Note that since we fix the mean of the distribution to be zero, increasing $\xi$ corresponds to the tail becoming lighter, i.e., $\bar{F}_{\sf N}(V) = 1 - F_{\sf N}(V)$ is decreasing as $\xi$ increases for a fixed $V \geq -1$. 
Increasing $\xi$ is analogous to the case of decreasing $p$ to zero for the two-point distribution of the niche type as defined in \eqref{eq:niche-two-point-distribution}. 
We compare the expected engagement and utility of {\sf APP} and {\sf DICE} in Figures \ref{fig:impact-heaviness-tail-delta-0} and \ref{fig:impact-heaviness-tail-delta-0.999} for $\delta = 0$ and $\delta = 0.999$ respectively. 
As the baseline, we consider {\sf APP} and the bar plots in Figures \ref{fig:impact-heaviness-tail-delta-0} and \ref{fig:impact-heaviness-tail-delta-0.999} demonstrate the relative change in engagement (in blue) and utility (in red) under {\sf DICE} compared to {\sf APP}. If a given metric (engagement or utility) is below/above the baseline {\sf APP}, it implies that {\sf DICE} loses/gains compared to the {\sf APP} with the percentage loss or gain depicted by the height of the bar plot.
The key insights are as follows:
\begin{enumerate}
    \item For a fixed discount factor $\delta \in [0,1)$, for larger tail parameter $\xi$ values (closer to 1), we observe misalignment between engagement and utility optimizing policies. In particular, we observe in Figures \ref{fig:impact-heaviness-tail-delta-0} and \ref{fig:impact-heaviness-tail-delta-0.999}, that ${\sf Eng}({\sf APP}) > {\sf Eng}({\sf DICE})$ and ${\sf Util}({\sf APP}) < {\sf Util}({\sf DICE})$ for large enough $\xi$. 
    \item As $\delta$ approaches $1$, i.e., as the platforms become more forward looking, we observe a stark asymmetry in the magnitude of the misalignment. In Figure \ref{fig:impact-heaviness-tail-delta-0.999}, we observe the there is little difference in the engagement obtained by {\sf DICE} and {\sf APP}. However, there is a significant enhancement in utility under {\sf DICE} compared to {\sf APP}.
\end{enumerate}

\begin{figure}[h]
    \centering
    \begin{tikzpicture}
    \pgfplotsset{
        scale only axis
    }
    \begin{axis}[
      axis y line*=left,
      axis x line*=bottom,
      xmin=0, xmax=1, 
      ymin=-0.15, ymax=0.31, 
      xlabel={$\xi$},
      ylabel={Rel. Change in Eng. and Util. under {\sf DICE} wrt {\sf APP}},
      yticklabel style={/pgf/number format/fixed,/pgf/number format/precision=3},
      xtick={0, 0.2, 0.4, 0.6, 0.8, 1},
      xticklabels = {$0$, $0.25$, $0.5$, $0.75$, $0.9$, $0.99$},
      ytick={-0.1, 0.1, 0.2, 0.3},
      yticklabels = {$-10\%$, $10\%$, $20\%$, $30\%$},
      extra y ticks={0},
      extra y tick labels = {\textcolor{purple}{\textbf{\large \sf APP}}},
      extra y tick style={grid=major, grid style={black, ultra thick}},
      ybar=-20pt, 
      bar width=20pt, 
      enlarge x limits=0.15, 
      legend style={at={(0.5,-0.15)},anchor=north,legend columns=-1} 
    ]

    \addplot+[draw=red, fill=red!90, mark=none] coordinates {
        (0, 0.88389598-1) (0.2, 0.91469701-1)  (0.6, 1.05696295-1)
        (0.8, 1.15476688-1) (1, 1.26733594-1)
    };
    \addlegendentry{Utility}

    \addplot+[draw=blue, fill=blue!90, mark=none] coordinates {
        (0, 0.94137585-1) (0.2, 0.93939899-1) (0.4, 0.93375514-1) (0.6, 0.92215147-1)
        (0.8,  0.90956446-1) (1, 0.89659493-1)
    };
    \addlegendentry{Engagement}
    \addplot+[draw = red, fill = red!90, mark = none] coordinates{
    (0.4, 0.96643944-1)
    };
    

    \draw[dashed, ultra thick, purple] (axis cs:0.5,-0.13) rectangle (axis cs:1.1,0.30);

    \node[text width=3.5cm] at (axis cs: 0.22,0.21) {\textcolor{purple}{misalignment b/w engagement and utility maximization}};
    
     \end{axis}
    
    \end{tikzpicture}
    \caption{Impact of \(\xi\) on utility and engagement for $\delta = 0$}
    \label{fig:impact-heaviness-tail-delta-0}
\end{figure}
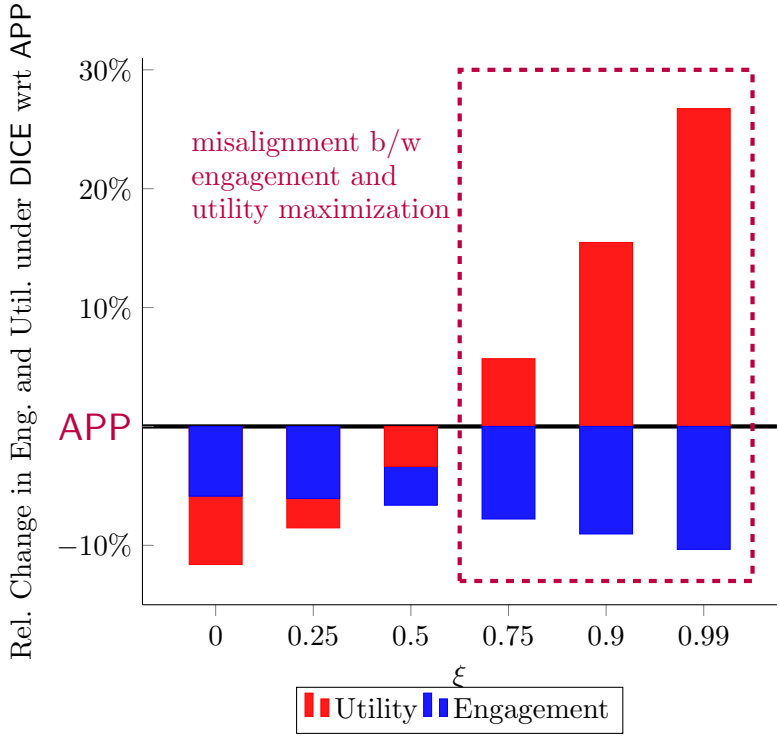

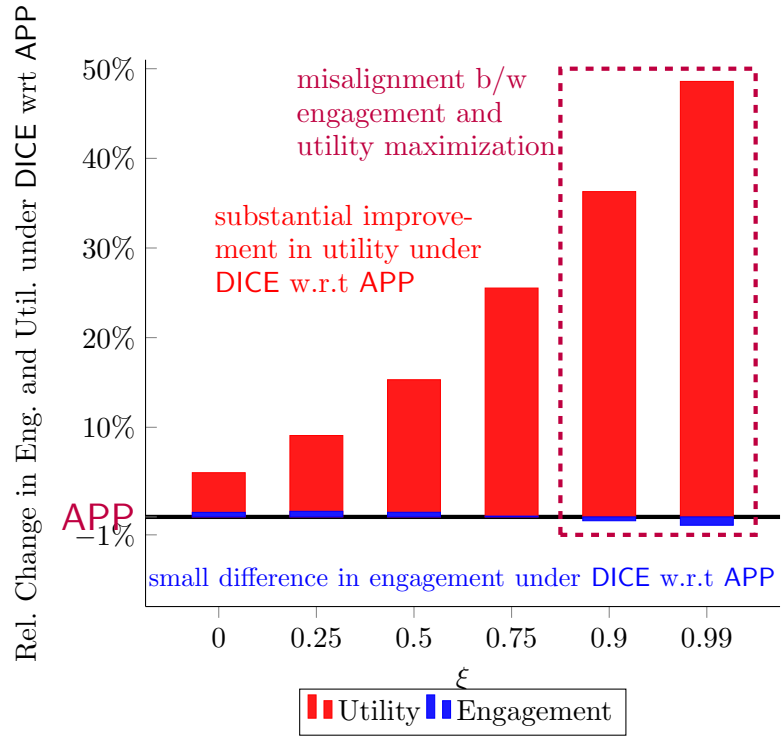
\begin{figure}[h]
    \centering
    \begin{tikzpicture}
    \pgfplotsset{
        scale only axis
    }

    \begin{axis}[
      axis y line*=left,
      axis x line*=bottom,
      xmin=0, xmax=1, 
      ymin=-0.1, ymax=0.51, 
      xlabel={$\xi$},
      ylabel={Rel. Change in Eng. and Util. under {\sf DICE} wrt {\sf APP}},
      yticklabel style={/pgf/number format/fixed,/pgf/number format/precision=3},
      xtick={0, 0.2, 0.4, 0.6, 0.8, 1},
      xticklabels = {$0$, $0.25$, $0.5$, $0.75$, $0.9$, $0.99$},
      ytick={-0.02, 0.1, 0.2, 0.3, 0.4, 0.5},
      yticklabels = {$-1\%$, $10\%$, $20\%$, $30\%$, $40\%$, $50\%$},
      extra y ticks={0},
      extra y tick labels = {\textcolor{purple}{\textbf{\large \sf APP}}},
      extra y tick style={grid=major, grid style={black, ultra thick}},
      ybar=-20pt, 
      bar width=20pt, 
      enlarge x limits=0.15, 
      legend style={at={(0.5,-0.15)},anchor=north,legend columns=-1} 
    ]

    \addplot+[draw=red, fill=red!90, mark=none] coordinates {
        (0, 1.04931316-1) (0.2, 1.09079237-1) (0.4, 1.15303408-1) (0.6, 1.25529535-1)
        (0.8, 1.36289228-1) (1, 1.4858012-1)
    };
    \addlegendentry{Utility}

    \addplot+[draw=blue, fill=blue!90, mark=none] coordinates {
        (0, 1.00504702-1) (0.2, 1.00622621-1) (0.4, 1.0051455-1) (0.6, 1.00088442-1)
        (0.8, 0.99575457-1) (1, 0.99071301-1)
    };
    \addlegendentry{Engagement}
    \node[text width=9cm] at (axis cs: 0.55,-0.07) {\small \textcolor{blue}{small difference in engagement under {\sf DICE} w.r.t {\sf APP}}};
    \node[text width=4cm] at (axis cs: 0.3,0.3) {\textcolor{red}{substantial improvement in utility under {\sf DICE} w.r.t {\sf APP}}};

    \draw[dashed, ultra thick, purple] (axis cs:0.7,-0.02) rectangle (axis cs:1.1,0.50);

    \node[text width=3.5cm] at (axis cs: 0.43,0.45) {\textcolor{purple}{misalignment b/w engagement and utility maximization}};
    
    \end{axis}
    
    \end{tikzpicture}
    \caption{Impact of \(\xi\) on utility and engagement for $\delta = 0.999$}
    \label{fig:impact-heaviness-tail-delta-0.999}
\end{figure}

\section{Conclusion}
\label{sec:conclusion}

In this work, we explore the extent of user utility loss when platforms use measurable but imperfect proxies like engagement to drive recommendations, and whether it is feasible to optimize for user utility, which is rarely measured. We studied a model where a recommendation system repeatedly suggests items to a user, adapting its selections over time. Our findings reveal a fundamental misalignment between policies that maximize engagement and those that maximize utility. We develop and study a utility-aware heuristic {\sf PEAR} which is able to achieve best of both worlds: near-optimal user utility and near-optimal engagement simultaneously. This highlights a stark asymmetry in the misalignment: substantial improvement in utility is achievable in comparison to the engagement-optimal policy {\sf APP} while sacrificing a minuscule amount on engagement. 
Moreover, we observe that some of these insights also carry over to the setting where the platform has no prior information about the utility distribution.
Overall, our research highlights that recommendation systems with the ability to recommend more than one item can facilitate {exploration with minimum reduction in engagement, allowing}
discovery of items with higher utility {and hence} 
leading to a significant enhancement in user utility.

Our model is intentionally simplified and does not encompass all aspects of contemporary recommendation systems. This simplicity allows us to focus on key tension between engagement and utility, however, our model may be expanded upon to study different facets of the recommendation systems. Amongst many open directions, our model can potentially be extended to (i) incorporate the setting with many niche types, (ii) model user satiation, where the utility at each time depends not only on the current consumption but past consumption as well, (iii) time-varying preferences and/or inconsistent preferences. We defer these extensions and other exciting open questions for future research.

\bibliographystyle{ACM-Reference-Format}
\bibliography{references}

\ECSwitch
\ECHead{Appendix}
The Appendix is organized as follows. Appendix \ref{app:useful-technical-result} provides a useful technical result characterizing the expected engagement and utility for a given recommendation set. Appendix \ref{app:proof-engagement-maximizing-policy-two-recommendations} provides the proof of Theorem \ref{thm:engagement-maximizing-policy-two-recommendations}. Appendix \ref{app:proof-utility-maximizing-policy-two-recommendations} provides the proof of Theorem \ref{thm:utility-maximizing-policy-two-recommednations}. Appendices \ref{app:proof-stopping-time-equivalence-lemma} and \ref{app:proof-g-limit-value} provide the proofs of helper lemmas used in the proof of Theorem \ref{thm:utility-maximizing-policy-two-recommednations}.

\section{Useful Technical Result}
\label{app:useful-technical-result}
The following proposition characterizes the probability of engagement and the expected utility of the user given the user is recommended items $\pi = \{i_1, i_2\}$.
\begin{proposition}
    \label{prop:engagement-utility-simplification}
    Given a recommendation $\pi = \{i_1, i_2\}$, the expected engagement and utility is given as
    \begin{align*}
        \mathbb{E}\left[ \mathbbm{1}\{c(\pi) \neq \emptyset \}\right] &= \mathbb{P}\left( c(\pi) \neq \emptyset \right) = \frac{e^{V_{\tau(i_1)}} + e^{ V_{\tau(i_2)}}}{1 + e^{V_{\tau(i_1)}} + e^{V_{\tau(i_2)}}} \\
        \mathbb{E}\left[ \max_{j \in \{i_1, i_2, \emptyset\}} u_j\right] &= \ln\left(1 + e^{V_{\tau(i_1)}} + e^{V_{\tau(i_2)}}\right)
    \end{align*}
\end{proposition}
For a proof of Proposition \ref{prop:engagement-utility-simplification}, refer to \citep[Chapter 2]{anderson1992discrete}.

\section{Proof of Theorem \ref{thm:engagement-maximizing-policy-two-recommendations}}
\label{app:proof-engagement-maximizing-policy-two-recommendations}
\proof{Proof.}
    We fix $\delta \in [0,1)$ and the attraction parameter of the popular type $V_{\sf P} \in \mathbb{R}_{+}$. The expected engagement under {\sf APP} is given as 
    \begin{align}
        \label{eq:always-popular-engagement}
        {\sf Eng}({\sf APP}) &= {\frac{2e^{V_{\sf P}}}{1 + 2e^{V_{\sf P}}}} + {\frac{\delta}{1 - \delta} \cdot \frac{2e^{V_{\sf P}}}{1 + 2e^{V_{\sf P}}}}
    \end{align}
    Let {\sf E-OPT} denote the engagement optimal policy in the class of all online policies $\Pi$. Consider the following exploration-based policy denoted {\sf Diverse-then-Optimal (DO)} -- it recommends one popular and one niche type at the first time step and from the second step onwards, implements {\sf E-OPT}. The expected engagement under {\sf DO} is given as
    \begin{align*}
        {\sf Eng}({\sf DO}) &= (1 - p) \cdot \frac{e^{V_{\sf P}} + e^{-1}}{1 + e^{V_{\sf P}} + e^{-1}} + p \cdot \frac{e^{V_{\sf P}} + e^{(1 - p) / p}}{1 + e^{V_{\sf P}} + e^{(1 - p) / p}} + \frac{\delta}{1 - \delta} \cdot {\sf Eng}({\sf E-OPT}) \\
        &\leq (1 - p) \cdot \frac{e^{V_{\sf P}} + e^{-1}}{1 + e^{V_{\sf P}} + e^{-1}} + p \cdot \frac{e^{V_{\sf P}} + e^{(1 - p) / p}}{1 + e^{V_{\sf P}} + e^{(1 - p) / p}} + \frac{\delta}{1 - \delta} \left[ p \cdot \frac{2e^{(1 - p) / p}}{1 + 2e^{(1 - p) / p}} + (1 - p) \cdot \frac{2e^{V_{\sf P}}}{1 + 2 e^{V_{\sf P}}} \right]
    \end{align*}
    where the inequality follows from the fact that the engagement under the optimal online policy is bounded above by the engagement of an oracle who knows whether a user prefers niche to popular type and recommends the preferred type. Then we have that 
    \begin{align*}
        \lim_{p \to 0} \left[{\sf Eng}({\sf APP}) - {\sf Eng}({\sf DO}) \right] = \frac{2e^{V_{\sf P}}}{1 + 2e^{V_{\sf P}}} - \frac{e^{V_{\sf P}} + e^{-1}}{1 + e^{V_{\sf P}} + e^{-1}} > 0
    \end{align*}
    Hence the limit $p \to 0$, we have that {\sf APP} does better than the best exploration-driven policy {\sf DO} and hence it is optimal.
    The expected utility under {\sf APP} is $\frac{1}{1 - \delta} \cdot \ln(1 + 2e^{V_{\sf P}})$ which follows as a corollary of Proposition \ref{prop:engagement-utility-simplification}. \QEDA
\endproof

\section{Proof of Theorem \ref{thm:utility-maximizing-policy-two-recommednations}}
\label{app:proof-utility-maximizing-policy-two-recommendations}
\proof{Proof.}
    We begin by defining some quantities. Fix a discount factor $\delta \in [0,1)$ and the parameter $p$ from \eqref{eq:niche-two-point-distribution}. Define the following
    \begin{align}
        \label{eq:constants}
        \rho_1 &\triangleq \frac{e^{(1 - p) / p}}{1 + e^{V_{\sf P}} + e^{(1 - p) / p}}, \ \ \ \ \ \ \rho_2 \triangleq \frac{e^{-1}}{1 + e^{V_{\sf P}} + e^{-1}}, \ \ \ \
        c &\triangleq \frac{\ln((1 - \rho_2) / (1 - \rho_1))}{\ln(\rho_1 / \rho_2) + \ln((1 - \rho_2) / (1 - \rho_1))}
    \end{align}
    For a $\rho, x \in [0,1]$, define the following random variable
    \begin{align}
        \label{eq:random-walk}
        X_{k}(\rho, x) = \begin{cases}
            1 - x,  & \quad \text{ with probability } \rho \\
            -x, & \quad \text{ with probability } 1 - \rho
        \end{cases}
    \end{align}
    Let $S_{n}(\rho, x) = \sum_{k = 1}^n X_{k}(\rho, x)$ denote the $n$-th partial sum of the random walk described by $X_k(\rho, x)$.
    Define $N(\rho, x)$ to be the first time the partial sum goes below zero, i.e.,
    \begin{align}
        \label{eq:stopping-time}
        N(\rho, x) &= \inf\{n : S_{n}(\rho, x) < 0\}
    \end{align}
    {\sf PEAR} recommends one item of popular type and another item of niche type while the posterior belief (denoted as $p_t$) on $V_{\sf N} = (1 - p) / p$ is greater than or equal to $p$. If the posterior $p_t < p$, then {\sf PEAR} switches to showing both items of the popular type. The time at each the policy switches from showing one popular and one niche type to both popular type is given by the random variable $N(\rho,c)$ defined in \eqref{eq:stopping-time} for different values of $\rho$ and this is characterized in the following lemma.
    \begin{lemma}
        \label{lem:stopping-time-equivalence}
        Consider $\rho_1,\rho_2$ and $c$ defined in \eqref{eq:constants}. If  $V_{\sf N} = (1 - p)/ p$, then the first time {\sf PEAR} recommends both the items of popular type is $N(\rho_1, c)$. Analogously, if $V_{\sf N} = -1$, then the first time {\sf PEAR} recommends both the items of the popular type is $N(\rho_2, c)$. 
    \end{lemma}
    We defer the proof of Lemma \ref{lem:stopping-time-equivalence} to Appendix \ref{app:proof-stopping-time-equivalence-lemma}. Lemma \ref{lem:stopping-time-equivalence} implies that from $t = 0$ to $t = N(\rho,c) - 1$, {\sf PEAR} recommends one popular and one niche type item and for $t \geq N(\rho,c)$, {\sf PEAR} recommends both items of the popular type. Hence the expected engagement under {\sf PEAR} is given as 
    \begin{align*}
        {\sf Eng}({\sf PEAR}) &= p \cdot \mathbb{E}\left[  \beta_1 \sum_{t = 0}^{N(\rho_1, c) - 1}\delta^{t} + \lambda \sum_{t = N(\rho_1, c)}^\infty \delta^t \right] + (1 - p) \cdot \mathbb{E}\left[\beta_2 \sum_{t = 0}^{N(\rho_2, c) - 1} \delta^t + \lambda \sum_{t = N(\rho_2, c)}^\infty \delta^t   \right] \\
        &= \frac{p}{1 - \delta} \cdot \left( \beta_1 g(\delta, \rho_1, c) + \lambda (1 - g(\delta, \rho_1, c)) \right)  + \frac{1 - p}{1 - \delta} \cdot \left( \beta_2 g(\delta, \rho_2, c) +  \lambda (1 - g(\delta, \rho_2, c))  \right),
    \end{align*}
    where $\beta_1 = \frac{e^{V_{\sf P}} + e^{(1 - p) /p}}{1 + e^{V_{\sf P}} + e^{(1 - p) / p}}, \beta_2 = \frac{e^{V_{\sf P}} + e^{-1}}{1 + e^{V_{\sf P}} + e^{-1}},  \lambda = \frac{2e^{V_{\sf P}}}{1 + 2e^{V_{\sf P}}}$ and $g(\delta, \rho, x)$ is defined as follows
    \begin{align}
        \label{eq:definition-g}
        g(\delta, \rho, x) &\triangleq \mathbb{E}\left[ 1 - \delta^{N(\rho,x )}\right].
    \end{align}
    Similarly, the expected utility under {\sf PEAR} is given as 
    \begin{align*}
        {\sf Util}({\sf PEAR}) &= p \cdot \mathbb{E}\left[  \Psi_1 \sum_{t = 0}^{N(\rho_1, c) - 1}\delta^{t} + \Lambda \sum_{t = N(\rho_1, c)}^\infty \delta^t \right]  + (1 - p) \cdot \mathbb{E}\left[\Psi_2 \sum_{t = 0}^{N(\rho_2, c) - 1} \delta^t + \Lambda \sum_{t = N(\rho_2, c)}^\infty \delta^t   \right] \\
        &= \frac{p}{1 - \delta} \cdot \left( \Psi_1 g(\delta, \rho_1, c) + \Lambda (1 - g(\delta, \rho_1, c)) \right) + \frac{1 - p}{1 - \delta} \cdot \left( \Psi_2 g(\delta, \rho_2, c) +  \Lambda (1 - g(\delta, \rho_2, c))  \right),
    \end{align*}
    where $\Psi_1 = \ln\left( 1 + e^{V_{\sf P}} + e^{(1 - p) / p}\right), \Psi_2 = \ln\left( 1 + e^{V_{\sf P}} + e^{-1}\right)$ and $\Lambda = \ln(1 + 2e^{V_{\sf P}})$. Finally, we are interested in the limit $p \to 0$. Note that $g(\delta, \rho_1, c)$ and $g(\delta, \rho_2, c)$ are also function of $p$. The following lemma characterize the limiting value of $g(\delta, \rho_1, c)$ and $g(\delta, \rho_2, c)$ when $p \to 0$. We defer the proof of Lemma \ref{lem:g-limit-value} to Appendix \ref{app:proof-g-limit-value}.
    \begin{lemma}
        \label{lem:g-limit-value}
        Fix $\delta \in [0,1)$ and consider $\rho_1, \rho_2$ and $c$ defined in \eqref{eq:constants}. Then $\lim_{p \to 0} g(\delta, \rho_1, c) = 1$ and $\lim_{p \to 0} g(\delta, \rho_2, c) = \frac{1 - \delta}{1 - \delta \rho_2}$.
    \end{lemma}
    In the limit $p \to 0$, the expected engagement and utility is
    \begin{align*}
        \lim_{p \to 0} {\sf Eng}({\sf PEAR}) &= \frac{1}{1 - \delta} \left(\beta_2 \cdot \frac{1 - \delta}{1 - \delta \rho_2} + \lambda \cdot \frac{\delta(1 - \rho_2)}{1 - \delta \rho_2} \right) \\
        \lim_{p \to 0} {\sf Util}({\sf PEAR}) &= \frac{1}{1 - \delta}\left(1 + \Psi_2 \cdot \frac{1 - \delta}{1 - \delta \rho_2} + \Lambda \cdot \frac{\delta(1 - \rho_2)}{1 - \delta \rho_2} \right)
    \end{align*}
    This concludes the proof. \QEDA
\endproof

\section{Proof of Helper Lemmas}
In this section, we provide the proof of some helper lemmas used in the proof of Theorem \ref{thm:utility-maximizing-policy-two-recommednations}.
\subsection{Proof of Lemma \ref{lem:stopping-time-equivalence}}
\label{app:proof-stopping-time-equivalence-lemma}
\proof{Proof.}
    Note that {\sf PEAR} switches to showing both the items of the popular type whenever $p_{t} < p$ where $p_t$ is the posterior belief of $V_{\sf N} = p / (1 - p)$. Note that $p_t$ depends on the (random) number of successes $S$ and failures $F$ observed till time $t$, where if we count the user choosing the niche item as success and the user not choosing the niche item (i.e. either choosing the popular item or the outside option) as failure. We have that the following are equivalent
    \begin{align*}
        p_t < p  &\stackrel{(a)}\iff \frac{1}{p} < \frac{1 - p}{p} \frac{\rho_2^{S} (1 - \rho_2)^{F}}{\rho_1^{S}(1 - \rho_1)^{F}} + 1 \\
        &\stackrel{(b)}\iff \rho_1^{S}(1 - \rho_1)^{F} < \rho_2^{S}(1 - \rho_2)^{F} \\
        &\stackrel{(c)}\iff S \left( \frac{\ln\left(\frac{\rho_1}{\rho_2}\right)}{\ln\left(\frac{\rho_1}{\rho_2}\right) + \ln\left(\frac{1 - \rho_2}{1 - \rho_1}\right)} \right) + F \left( -\frac{\ln\left(\frac{1 - \rho_2}{1 - \rho_1}\right)}{\ln\left(\frac{\rho_1}{\rho_2}\right) + \ln\left(\frac{1 - \rho_2}{1 - \rho_1}\right)} \right) < 0 \\
        &\stackrel{(d)}\iff S(1 - c) + F (-c) < 0,
    \end{align*}
    where $(a)$ follows from the definition of $p_t$, $(b)$ and $(c)$ follows from algebraic manipulations and $(d)$ follows from the definition of $c$ in \eqref{eq:constants}. Note that the equation $S(1 - c) + F (-c)$ corresponds to the $S + F$-th partial sum of a random walk with steps of size $1 - c$ or $-c$, which is the same random walk as described in \eqref{eq:random-walk}. Therefore $N(\rho_1, c)$ and $N(\rho_2, c)$ correspond to the stopping time when $V_{\sf N} = p/(1 - p)$ and $V_{\sf N} = -1$. \QEDA
\endproof

\subsection{Proof of Lemma \ref{lem:g-limit-value}}
\label{app:proof-g-limit-value}
\proof{Proof.}
    Define $T(\rho, c) = \inf\{n : X_n(\rho, c) = -c\}$ for $\rho \in \{ \rho_1, \rho_2\}$ and $M_0 = \frac{\ln\left((1 - \rho_2)/ (1 - \rho_1)\right)}{\ln\left({\rho_1} / { \rho_2 } \right)} = \Theta(1 /p)$. Recall the random variable $N(\rho,x)$ is defined in \eqref{eq:stopping-time}. Now for all $k \leq M_0$, we have that the following events are equivalent,
    \begin{align}
        \label{eq:event-equivalence-rho}
        \{T(\rho, c) = k\} \equiv \{N(\rho, c) = k\} \ \ \text{for } \rho \in \{\rho_1, \rho_2\}
    \end{align}
    We first show that $\lim_{p \to 0} g(\delta, \rho_1, c) = 1$. Since $\delta \in (0,1]$, we trivially have that $g(\delta, \rho_1, c) \leq 1$. Next we will lower bound $g(\delta, \rho_1, c)$. 
    \begin{align*}
        g(\delta, \rho_1, c) &\stackrel{(a)}= 1  - \sum_{k = 1}^\infty \delta^k \mathbb{P}(N(\rho_1, c) = k) \\ 
        &\stackrel{(b)}= 1 - \sum_{k = 1}^{M_0} \delta^k \mathbb{P}(T(\rho_1, c) = k) - \sum_{k = M_0 + 1}^\infty \delta^k \mathbb{P}(N(\rho_1,k) = k)\\
        &\stackrel{(c)}\geq 1 - \frac{1 - \rho_1}{\rho_1} \sum_{k = 1}^{M_0} \delta^k \rho_1^k - \delta^{M_0 + 1}\\ 
        &\stackrel{(d)}= 1 - \frac{\delta  (1 - \rho_1)  (1 - (\delta \rho_1)^{M_0})}{1 - \delta \rho_1} - \delta^{M_0 + 1} \\
        &\stackrel{(e)}\geq 1 - e^{-1/p}\frac{e\delta (1 + e^{V_{\sf P}})}{1 - \delta} - \delta^{\Theta(1 / p)} 
    \end{align*}
    where $(a)$ follows from the definition of $g(\delta, \rho, x)$ in \eqref{eq:definition-g}, $(b)$ follows from \eqref{eq:event-equivalence-rho} for $\rho = \rho_1$, $(c)$ follows from the fact that $\mathbb{P}(T(\rho_1, c) = k) = (1 - \rho_1)\rho_1^{k - 1}$ and the fact that $\delta^{M_0} \geq \delta^k$ for all $k \geq M_0$, $(d)$ follows from sum of geometric series, $(e)$ follows from the fact that $1 - \rho_1 \leq (1 + e^{V_{\sf P}})e^{-1/p + 1}$, $1 - (\delta \rho_1)^{M_{0}} \leq 1$ and $1 - \delta \rho_1 \geq 1 - \delta$. Taking the limit of $p \to 0$ for a fixed $\delta \in [0,1)$ and $V_{\sf P} \in \mathbb{R}_{+}$, we have that $$\lim_{p \to 0} g(\delta, \rho_1, c) \geq \lim_{p \to 0} \left(1 - e^{-1/p}\frac{e\delta (1 + e^{V_{\sf P}})}{1 - \delta} - \delta^{\Theta(1/p)} \right) = 1$$

    Next we want that $\lim_{p \to 0} g(\delta, \rho_2, c) = \frac{1 - \delta}{1 - \delta \rho_2}$. 
    In particular, we will show that 
    \begin{align*}
        \frac{1 - \delta}{1 - \delta \rho_2} + \frac{e(1 + e^{V_{\sf P}})(\delta \rho_2)^{M_0 + 1}}{1 - \delta \rho_2} - \delta^{M_0 + 1} \leq g(\delta, \rho_2, c) \leq \frac{1 - \delta}{1 - \delta \rho_2} + \frac{e(1 + e^{V_{\sf P}})(\delta \rho_2)^{M_0 + 1}}{1 - \delta \rho_2}
    \end{align*}
    We will begin with the upper bound. 
    \begin{align*}
        g(\delta, \rho_2, c) &\stackrel{(a)}= 1 - \sum_{k = 1}^\infty \delta^k \mathbb{P}(N(\rho_2, c) = k), \\
        &\stackrel{(b)}\leq 1 - \sum_{k = 1}^{M_0} \delta^k \mathbb{P}(T(\rho_2, c) = k), \\
        &\stackrel{(c)}= 1 - \frac{1 - \rho_2}{\rho_2} \sum_{k = 1}^{M_0} \delta^k \rho_2^k, \\ 
        &\stackrel{(d)}= 1 - e (1 + e^{V_{\sf P}}) \frac{\delta \rho_2 (1 - (\delta \rho_2)^{M_0})}{1 - \delta \rho_2}, \\
        &\stackrel{(e)}= \frac{1 - \delta \rho_2 - e(1 + e^{V_{\sf P}})\delta \rho_2 + e(1 + e^{V_{\sf P}})(\delta \rho_2)^{M_0 + 1}}{1 - \delta \rho_2}, \\
        &\stackrel{(f)}= \frac{1 - \delta}{1 - \delta \rho_2} + \frac{e(1 + e^{V_{\sf P}})(\delta \rho_2)^{M_0 + 1}}{1 - \delta \rho_2}, 
    \end{align*}
    where $(a)$ follows from the definition of $g(\delta, \rho, x)$ in \eqref{eq:definition-g}, $(b)$ follows from \eqref{eq:event-equivalence-rho} for $\rho = \rho_2$, $(c)$ follows from the fact that $\mathbb{P}(T(\rho_2, c) = k) = (1 - \rho_2)\rho_2^{k - 1}$, $(d)$ follows from sum of the geometric series, $(e)$ follows trivially, $(f)$ follows from the fact that $e(1 + e^{V_{\sf P}})\delta \rho_2 = (1 - \rho_2)$. The lower bound on $g(\delta, \rho_2,c)$ follows using a similar line of argument. Note that as $p \to 0$, since $\delta \rho_2 < 1$ and $\delta < 1$, we have that $(\delta \rho_2)^{M_0 + 1} \to 0$ and $\delta^{M_0 + 1} \to 0$ and together we have that $\lim_{p \to 0} g(\delta, \rho_2, c) = \frac{1 - \delta}{1 - \delta \rho_2}$. \QEDA
\endproof

\end{document}